**Proteomics Standards Initiative's ProForma 2.0: Unifying the Encoding of Proteoforms and Peptidoforms**

Richard D. LeDuc[1,2*], Eric W. Deutsch[3], Pierre-Alain Binz[4], Ryan T. Fellers[1], Anthony J. Cesnik[5,6,7], Joshua A. Klein[8], Tim Van Den Bossche[9,10], Ralf Gabriels[9,10], Arshika Yalavarthi[1], Yasset Perez-Riverol[11], Jeremy Carver[12,13,14], Wout Bittremieux[12,14], Shin Kawano[15,16], Benjamin Pullman[12,13], Nuno Bandeira[12,13,14], Neil L. Kelleher[1], Paul M. Thomas[1,17], Juan Antonio Vizcaíno[11,*]

[1] National Resource for Translational and Developmental Proteomics, Northwestern University, Evanston, IL, 60611, USA.

[2] Present Address: Children's Hospital Research Institute of Manitoba, John Buhler Research Centre, 715 McDermot Avenue, Winnipeg, Manitoba, R3E 3P4.

[3] Institute for Systems Biology, Seattle WA 98109, USA.

[4] Clinical Chemistry Service, Lausanne University Hospital, 1011 Lausanne, Switzerland, 1007.

[5] Department of Genetics, Stanford University, Stanford, CA 94305, USA.

[6] Chan Zuckerberg Biohub, 499 Illinois St, San Francisco, CA 94158, USA.

[7] SciLifeLab, School of Engineering Sciences in Chemistry Biotechnology and Health, KTH – Royal Institute of Technology, SE-171 21 Solna, Stockholm, Sweden, 113 51.

[8] Program for Bioinformatics, Boston University, Boston, MA 02215, USA.

[9] VIB – UGent Center for Medical Biotechnology, VIB, Technologiepark 75 - FSVM II, 9052 Ghent, Belgium.

[10] Department of Biomolecular Medicine, Faculty of Medicine and Health Sciences, Ghent University, 9000 Ghent, Belgium.

[11] European Molecular Biology Laboratory, EMBL-European Bioinformatics Institute (EMBL-EBI), Hinxton, Cambridge, CB10 1SD, United Kingdom.

[12] Center for Computational Mass Spectrometry, University of California, San Diego (UCSD), La Jolla, CA 92093, USA.

[13] Dept. Computer Science and Engineering, University of California, San Diego (UCSD), La Jolla, CA 92093, USA.

[14] Skaggs School of Pharmacy and Pharmaceutical Sciences, University of California, San Diego (UCSD), La Jolla, CA 92093, USA.

[15] Toyama University of International Studies, Toyama. 930-1292 Toyama, Higashikuromaki, 6 5-1, Japan

[16] Database Center for Life Science, Joint Support-Center for Data Science Research, Research Organization of Information and Systems, Kashiwa, Chiba 277-0871, Japan.

[17] Present Address: AbbVie, Inc., 1401 Sheridan Rd, North Chicago, IL 60064 USA.

[*] Corresponding authors: Richard D. LeDuc (RLeduc@chrim.ca) & Juan Antonio Vizcaíno (juan@ebi.ac.uk)

**Abstract**

It is important for the proteomics community to have a standardized manner to represent all possible variations of a protein or peptide primary sequence, including natural, chemically-induced and artifactual modifications. The Human Proteome Organization (HUPO) Proteomics Standards Initiative (PSI) in collaboration with several members of the Consortium for Top-Down Proteomics (CTDP) has developed a standard notation called ProForma 2.0, which is a substantial extension of the original ProForma notation developed by the CTDP. ProForma 2.0 aims to unify the representation of proteoforms and peptidoforms.

ProForma 2.0 supports use cases needed for bottom-up and middle-/top-down proteomics approaches and allows the encoding of highly modified proteins and peptides using a human- and machine-readable string. ProForma 2.0 can be used to represent protein modifications in a specified or ambiguous location, designated by mass shifts, chemical formulas, or controlled vocabulary terms, including cross-links (natural and chemical), and atomic isotopes. Notational conventions are based on public controlled vocabularies and ontologies. The most up-to-date full specification document and information about software implementations are available at http://psidev.info/proforma.

## Introduction

Protein and peptide sequences are usually represented by a string of amino acids using the well-known one-letter code that was first introduced by the International Union of Pure and Applied Chemistry (IUPAC) in 1972.[1] The linear arrangement of the amino acids is customarily written from the *N*-terminus to the *C*-terminus. However, there is no clear consensus about how to represent amino acid modifications, which can be natural [e.g., biologically-relevant post-translational modifications (PTMs)], chemically-induced (including, for example, reduction/alkylation and addition of tags for quantitative analysis) or artifactual as a consequence of sample preparation (such as oxidation and deamidation).

The terms "proteoform"[2] and "peptidoform,"[3] are used for the specific "form" or "entity" of a given protein or peptide that results from the combination of the amino acid sequence and modification(s) at specific amino acid positions. Multiple proteoforms can be derived from the same gene. For example, if a protein has two sites that can potentially be phosphorylated, there are four possible proteoforms: the unmodified form represented by the primary sequence, and the forms with phosphorylation on the first site, the second site, and both sites. Each of these are distinct proteoforms, but only the first proteoform, the unmodified variant, can be written using the IUPAC notation. In the absence of a recognized standard notation, there is no consistency in the way modified proteins and peptides are designated. This can not only lead to confusion in scientific publications and presentations, but it is also a major dilemma for developers of proteomics software and resources to decide what notation(s) to use for data input and output. This is applicable to widely-used protein-centric database resources such as UniProtKB (UniProt Knowledge-Base),[4] ProteomeXchange proteomics resources,[5] the Protein Data Bank (PDB),[6] Reactome[7] and IntAct,[8] among many others. This has led to the development of multiple different notational formats by various groups.

In order to make peptide and protein data more *findable*, *accessible*, *interoperable* and *reusable* (FAIR),[9] there needs to be a single IUPAC-compatible notational standard to encode modified protein and peptide sequences. In 2018, the Consortium for Top-Down Proteomics (CTDP) introduced the ProForma notation,[10] which answered the immediate needs of the Consortium by creating a standardized method for designating a proteoform. It contained seven rules to denote both the primary structure of a proteoform and most of the commonly-observed PTMs and artifactual modifications, using nomenclature from five ontologies and controlled vocabularies (CVs). In general, CVs are minimally structured lists of terms and definitions, while ontologies encode the full hierarchical relationship structure among the terms[11].

However, this notational system was not sufficient to meet the needs of the broader proteomics community and protein data resources because some important use cases were not supported. In particular, the first ProForma version did not address issues such as ambiguity in either the order of the amino acid sequence or modification site localization, and did not support cross-links (natural or chemically-induced), among many others. For proteoform and peptidoform designations to be FAIR across the broader array of protein science data resources, these and numerous other notational issues needed to be addressed. Ideally, the same notational system should be usable for both bottom-up and middle-/top-down applications.

The Proteomics Standards Initiative (PSI) of the Human Proteome Organization (HUPO) develops and ratifies community-based data standards and CVs for the field of proteomics,[12] including mzML,[13] mzIdentML,[14] mzTab,[15] PSI-MOD,[16] PEFF (PSI Extended FASTA Format)[17] and more recently, the Universal Spectrum Identifier (USI)[18] and the sample metadata standard MAGE-TAB-Proteomics.[19] Each of these standards has been subjected to the PSI Document Process[20] which mandates three levels of review that must be completed before a proposed standard is ratified. In order to address the use cases needed for bottom-up and middle-/top-down approaches, members of the CTDP and HUPO-PSI worked together and devised an extended ProForma notation designed to meet the current and future needs for protein sequence data. In this article, we present an overview of the ProForma 2.0 notation, a brief description of its most salient features, and some example applications.

## Methods

### Development of ProForma 2.0

The development of ProForma 2.0 started in 2019. Since then, it was an open process via conference calls in addition to discussions at the annual PSI meetings and smaller workshops. The ProForma 2.0 specification document was submitted to the PSI Document Process for review, during which time external reviewers provided their feedback. The document was also made available for comments by the public, enabling broad input on the specifications. The final version of the ProForma 2.0 specification document is provided as Supplementary Document 1. Potential corrections to the document, up-to-date information on software implementations, and information on future versions of ProForma are available at http://psidev.info/proforma.

The main requirements considered during the development of the standard notation were:

1. It must be a string of characters that is human-readable, i.e. it should be suitable for display in a written document or in a presentation.

2. It must be unambiguously parsable by software (i.e., machine-parsable).

3. It must be able to support the encoding of amino acid sequences and their modifications (including natural, chemically-induced and artifactual).

4. It must be able to support the main use cases needed by the proteomics community as a whole, including bottom-up (focused on peptides/peptidoforms) and middle-/top-down (focused on proteins/proteoforms) applications.

5. It must be flexible enough to accommodate different styles of notations that are currently in common use.

6. It must be compatible with other existing PSI file formats.

7. It must be able to accommodate ambiguity in the position of a modified site.

8. It must be able to evolve so that new use cases can be added in the future.

Requirements 1 – 3 were included in the original ProForma 1.0 notation.[10] The essence of the fourth requirement was in the ProForma 1.0 notation, but the current version now includes support for bottom-up proteomics-specific entities, i.e. for peptidoforms, whereas the original exclusively defined the way to designate whole proteoform sequences. Requirements 4 – 8 are new in ProForma 2.0.

An essential requirement of ProForma 2.0 is that it should be able to represent peptidoforms and proteoforms in a consistent and reproducible way, taking into consideration the different strategies for designating protein modifications. Moreover, it must be able to be used jointly with USIs[18] to represent peptide spectrum matches (PSMs) and proteoform spectrum matches (PrSMs).

**Results**

**Data Format Description**

Here we provide a brief overview with examples of the main features of ProForma 2.0, while the full ProForma 2.0 specification document, as ratified by the PSI, provides exhaustive details on all aspects of the data format. ProForma 2.0 provides a standardized set of rules for describing the location and nature of all mass modifications on a proteoform or peptidoform. An example is shown in Figure 1. Using ProForma 2.0, there is a string of characters that linearly represents the peptidoform/proteoform primary structure, with allowance for some level of ambiguity, and the possibility to link peptide chains together, such as by cross-linking. ProForma 2.0 is not intended to represent secondary or higher-order structures. ProForma 2.0 can also be used to represent the molecular interpretation of a tandem mass spectrum. It should be noted that ProForma 2.0 is designed to describe a single, specific peptidoform or proteoform and not a collection of protein sequences or a listing of all potential mass modifications that may be found on them (i.e., a protein sequence search database). Other file formats such as PEFF[17] are better suited for this purpose.

When using the ProForma 2.0 notation for peptidoforms and proteoforms, amino acids are shown as is customary from left to right, *N*- to *C*-terminus, using IUPAC single letter identifiers. Modifications of this core set of amino acids are designated by a coded string of characters enclosed in square brackets after the letter of the modified residue. The modification string is represented by CV or ontology terms. The supported CVs/ontologies in ProForma 2.0 are PSI-MOD,[16] Unimod,[21] RESID,[22] XL-MOD (cross-linking; https://github.com/HUPO-PSI/xlmod-CV) and the Glycan Naming Ontology (GNO; glycans; https://www.ebi.ac.uk/ols/ontologies/gno).

ProForma 2.0 is case insensitive. This means that the notation is agnostic with regard to the use of uppercase or lowercase characters. However, different CVs and/or ontologies generally have their own specific policies for capitalization and representation of terms. It is, therefore, recommended that the capitalization specifications for each supported CV/ontology be used. It is also important to highlight that line breaks must not be used. There is currently no limit in maximum length since ProForma 2.0 can be used to represent both peptidoforms and proteoforms. Additionally, non-ASCII (American Standard Code for Information Interchange) characters are allowed since they may be included in the supported terms in the different CVs/ontologies.

A comparison of the features of ProForma 1.0 (finished in 2018) and 2.0 is shown in Table 1. At least 18 features were either added or expanded. Examples of ProForma 2.0 notations are provided in Table 2, along with the section number in the specification document (Supplementary Document 1) that contains the detailed description of each feature. Note that custom user-specific information may be added to ProForma 2.0 entities by means of using "Information tags." Additionally, in the 2.0 version, the use of "Information tags" is the only mechanism to add metadata for a ProForma entity.

**Levels of Compliance**

It is important to highlight that software that implements the ProForma 2.0 notation may not support all aspects of the specification. For example, a standard proteomics search engine that outputs the ProForma notation does not have to support the cross-linking part of the notation. We have, therefore, defined five levels of ProForma 2.0 compliance (listed below) in order to make adoption easier. Details can be found in the specification document (Supplementary Document 1, Appendix I).

Base Level ("Base-ProForma Compliant").

Level 2 ("Level 2-ProForma compliant").

Top-Down Extensions (Level 2-ProForma + top-down compliant).

Cross-Linking Extensions (Level 2-ProForma + cross-linking compliant).

Glycan Extensions (Level 2-ProForma + glycans compliant).

More than one of the extensions listed above (top-down, cross-linking and glycan) could be supported by the same software.

**Software Implementations**

ProForma 2.0 has already been implemented in some existing software. The CTDP has established an initial proteoform registry where experimentally verified proteoforms are assigned a unique PFR (ProteoForm Record) identifier (http://www.proteoform.org/api).[23] This identifier system is essential for enhancing interoperability between tools and databases that include proteoform data. The registry is based on an API (Application Programming Interface) that accepts ProForma 2.0 sequences, compares them to known proteoforms already stored in the registry, and returns a new PFR identifier, if the proteoform is new to the system. However, if the proteoform is already stored in the registry, a PFR identifier generated previously is returned. Then, ProForma 2.0 is needed as an input to the registry so that PFR identifiers can be provided.

There are currently four implementations of parsers and writers for ProForma 2.0, including the following:

1. A .NET version, as part of the Top-Down Software Development Kit (SDK) (https://github.com/topdownproteomics/sdk). This includes a lexer/parser with some additional proteoform validation functionality.

2. A Java port of the .NET reader and writer (https://github.com/NRTDP/proforma-java).

3. A Python version of a parser and writer, which is now part of the Pyteomics [24] framework (https://github.com/levitsky/pyteomics). Additional documentation is available here (https://pyteomics.readthedocs.io/en/latest/api/proforma.html).

4. The spectrum_utils Python package [25] includes a parser using a formal grammar to convert ProForma strings into abstract syntax trees (https://github.com/bittremieux/spectrum_utils/).

ProForma strings are also an optional part of the recently developed USI standard for representation of PSMs (see some examples at http://proteomecentral.proteomexchange.org/usi/). We expect that adoption of ProForma will increase broadly in the field, stimulated by its inclusion in widely-used bioinformatics resources such as those created by the CTDP, ProteomeXchange[5] and UniProtKB,[4] among others.

**Discussion and Conclusions**

ProForma 2.0 is a standard notation that is capable of supporting the needs of both the bottom-up and the top-down proteomics communities. Since peptidoforms and proteoforms are easily encoded in the ProForma 2.0 notation, it simplifies comparing the results of different search engines. This will greatly facilitate reuse of experimental data. We also anticipate that the ProForma 2.0 notation will expedite integration of bottom-up and middle-/top-down data, which is an active field of research.[26, 27] Moreover, the notation can be used as an input for the first version of the Proteoform Registry, which generates of unambiguous PFR identifiers for proteoform entities. Use of PFR identifiers is key to facilitate proteoform data interoperability between multiple tools and protein databases.

Proforma 2.0 has been developed as a joint effort between the PSI and the CTDP and will be actively maintained. Both organizations expect that this version 2.0 will not change for an extended period of time since it addresses most of the relevant use cases at the time of writing. However, additional use cases have already been envisioned and documented in the specification document (see Section 5, "Pending Issues - Future developments," in Supplementary Document 1). We expect that these extra features can be addressed in future versions, after the community has gained experience with the more common use cases included in version 2.0. The current list of known open issues includes: representation of cyclic peptides, representation of more complex scenarios where there is ambiguity in the localization of different glycans attached to the same amino acid sequence, support for rare amino acids which are not assigned to an accepted one-letter code, support the use of average masses in the notation, lipid modifications, support for molecular formulas, overlapping ranges of possible protein modification localizations, ambiguous cross-linker modification positions, representation of the distribution of different isotopes in the sequence, and the representation of sequences coming from non-MS-based proteomics approaches (e.g. peptide nanopores and Edman-based sequencing).

PSI standards are developed via an open process in which all interested individuals and groups are encouraged to participate. ProForma 2.0 has been developed by contributors from both the top-down and bottom-up proteomics subfields. This fusion provides the community with a standard that supports a diverse array of use cases and creates the potential for a substantially higher degree of software tool interoperability within the field than in the past. Although standards that are cooperatively developed inevitably take longer to complete than formats proposed by a single group, the resulting standards are more broadly applicable to many more use cases than those from independent initiatives. Broad participation is, therefore, essential for successful generation of future standards for the proteomics community. See https://www.topdownproteomics.org/ to become involved in the top-down proteomics activities of the CTDP and https://psidev.info/ for information about how to contribute to the PSI.

## Supporting Information

**Supplementary Document 1:** ProForma 2.0 specification document.

## Author Information

Corresponding authors:

Richard D. LeDuc (RLeduc@chrim.ca)

 National Resource for Translational and Developmental Proteomics, Northwestern University, Evanston, IL, 60611, USA.

 Present Address: Children's Hospital Research Institute of Manitoba, John Buhler Research Centre, 715 McDermot Avenue, Winnipeg, Manitoba, R3E 3P4.

Juan Antonio Vizcaíno (juan@ebi.ac.uk)

 European Molecular Biology Laboratory, EMBL-European Bioinformatics Institute (EMBL-EBI), Hinxton, Cambridge, CB10 1SD, United Kingdom.

## Notes

The authors declare no competing financial interest.

**Acknowledgements**

The authors would like to acknowledge all the individuals who contributed in various ways to the ProForma 2.0 specification, (see specification document). We would also like to thank the reviewers of the PSI specification document, namely Gloria Sheynkman and Erin Jeffery (University of Virginia) and Xiaowen Liu (Tulane University School of Medicine). We are also appreciative of the contributions made by the Executive Board of the CDTP. We are especially grateful to Susan Weintraub (University of Texas Health Science Center at San Antonio) for having edited the manuscript quite extensively, to make it more suitable for readers who are non-experts in bioinformatics.

This work was financially supported in part by the following: JAV, National Institutes of Health (R24GM127667), BBSRC (BB/S01781X/1), the EU H2020 project EPIC-XS (823839) and EMBL core funding; RDL, RTF, PMT and NLK, National Institute of Health (P41 GM108569), the Human Biomolecular Atlas Program (UH3 CA246635) and the National Library of Medicine (R21 LM013097); SK, the Database Integration Coordination Program from the National Bioscience Database Center, Japan Science and Technology Agency (18063028) and the Japan Society for the Promotion of Science KAKENHI (JP20H03245); AJC, the Knut and Alice Wallenberg Foundation (2016.0204) and the Swedish Research Council (2017-05327 to Emma Lundberg); TVDB,  the Research Foundation – Flanders (1S90918N); EWD, the National Institutes of Health (R24GM127667, R01GM087221, U19AG023122), and the National Science Foundation (DBI-1933311); NB, the National Institutes of Health (R24GM127667, and 1R01LM013115) and the National Science Foundation (ABI 1759980).

**Figure and Figure legend**

**Figure 1.** Representation of the same *N*-terminal segment (sharing the same amino acid sequence) of two hypothetical proteoforms using ProForma 2.0: the unmodified proteoform (top part of the figure) and one containing different protein modifications (lower part of the figure). The text coloration is only included here to improve clarity. The purple tag encodes the existence of an unlocalized phosphorylation event somewhere on the proteoform. The keyword "Phospho" is from Unimod and can be used without additional clarification. The brown tag is a reference to an *N*-terminal modification using the term "Acetyl" from Unimod. A 174.3-Da mass shift on the arginine is also indicated.

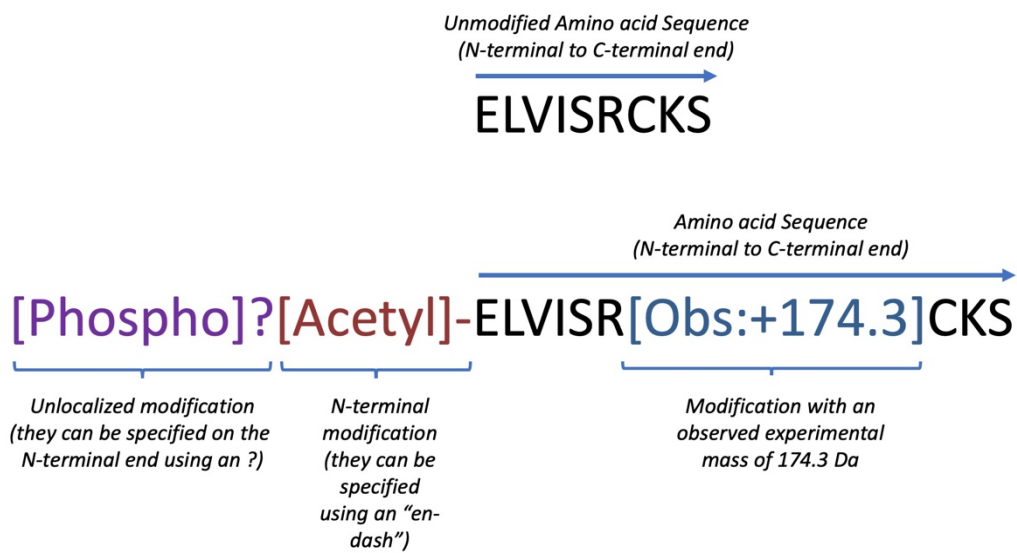

**Table 1.** Comparison of the supported features of ProForma 1.0 and 2.0.

| Feature | ProForma 1.0 | ProForma 2.0 |
|---|---|---|
| Protein modifications designated by CV/ontology names and accession numbers | ✓ | ✓ |
| Representation of glycan composition | ✓ | ✓ |
| *N*-terminal and *C*-terminal modifications | ✓ | ✓ |
| Delta mass notation for modifications | ✓ | ✓ |
| Information tag | ✓ | ✓ |
| Joint representation of experimental data and interpretation | ✓ | ✓ |
| NEW Support for elemental formulas | Limited | ✓ |
| NEW Representation of isotopes | Limited | ✓ |
| NEW Cross-link notation | X | ✓ |
| NEW Representation of inter-chain cross-links | X | ✓ |
| NEW Representation of disulfide linkages | X | ✓ |
| NEW Representation of glycans with GNO ontology as CV | X | ✓ |
| NEW Specifying a gap of known mass | X | ✓ |
| NEW Labile modifications | X | ✓ |
| NEW Unknown modification position | X | ✓ |
| NEW Possible set of modification positions | X | ✓ |
| NEW Representing ranges of positions for the modifications | X | ✓ |
| NEW Modification position preference and localization scores | X | ✓ |
| NEW Scoring for ranges of positions for a modification | X | ✓ |
| NEW Fixed protein modifications | X | ✓ |
| NEW Ambiguity in the order of amino acid sequences | X | ✓ |
| NEW Representation of ion charges and more than one peptidoform per spectrum | X | ✓ |
| NEW Representation of branched peptides | X | ✓ |
| NEW Representation of ambiguity in the order of the amino acid sequence | X | ✓ |

**Table 2.** Examples of ProForma 2.0 notations demonstrating the various features of the specification. For each feature listed in the first column, there is a representative example in the second column showing the encoding. The "Section" column provides the location in the PSI specification document where the feature is explained in detail (Supplementary Document 1).

| Feature | Example | Section |
|---------|---------|---------|
| CV/ontology modification names | EM[Oxidation]EVEES[Phospho]PEK | 4.2.1 |
| CV/ontology protein modification accession numbers | EM[MOD:00719]EVEES[MOD:00046]PEK | 4.2.2 |
| Cross-link within the same peptide | EMEVTK[XLMOD:02001#XL1]SESPEK[#XL1] | 4.2.3.1 |
| Inter-chain cross-links | SEK[XLMOD:02001#XL1]UENCE//EMEVTK[#XL1]SESPEK | 4.2.3.2 |
| Disulfide linkages | EVTSEKC[MOD:00034#XL1]LEMSC[#XL1]EFD | 4.2.3.3 |
| Branched peptides | ETFGD[MOD:00093#BRANCH]//R[#BRANCH]ATER | 4.2.4 |
| Glycans using the GNO ontology as CV | NEEYN[GNO:G59626AS]K | 4.2.5 |
| Delta mass notation for modifications | EM[+15.9949]EVEES[+79.9663]PEK | 4.2.6 |
| Specifying a gap of known mass | RTAAX[+367.0537]WT | 4.2.7 |
| Support for elemental formulas | SEQUEN[Formula:C12H20O2]CE | 4.2.8 |
| Glycan composition | SEQUEN[Glycan:HexNAc1Hex2]CE | 4.2.9 |
| *N*-terminal and *C*-terminal modifications | [iTRAQ4plex]-EMEVNESPEK | 4.3.1 |
| Labile modifications | {Glycan:Hex}EMEVNESPEK | 4.3.2 |
| Unknown modification position | [Phospho]?EMEVTSESPEK | 4.4.1 |
| Possible set of modification positions | EMEVT[#g1]S[#g1]ES[Phospho#g1]PEK | 4.4.2 |
| Ranges of positions for the modifications | PROT(ESFRMS)[+19.0523]ISK | 4.4.3 |
| Modification position preference and localization scores | EMEVT[#g1(0.01)]S[#g1(0.09)]ES[Phospho#g1(0.90)]PEK | 4.4.4 |
| Scoring for ranges of positions for a modification | PROT(ESFRMS)[+19.0523#g1(0.01)]ISK[#g1(0.99)] | 4.4.5 |
| Isotopes | <13C>ATPEILTVNSIGQLK | 4.6.1 |
| Fixed protein modifications | <[MOD:01090]@C>ATPEILTCNSIGCLK | 4.6.2 |
| Ambiguity in the order of the amino acid sequence | (?DQ)NGTWEMESNENFEGYMK | 4.7 |
| Information tag | ELVIS[Phospho|INFO:newly discovered]K | 4.8 |
| Joint representation of experimental data and interpretation | ELVIS[Phospho|Obs:+79.978]K | 4.9 |
| Representation of ion charges | EMEVEESPEK/2 | 7.1 |
| Multiple peptidoforms assigned to chimeric spectra | EMEVEESPEK/2+ELVISLIVER/3 | 7.1 |

**Abbreviations**

API: Application Programming Interface

ASCII: American Standard Code for Information Interchange

CTDP: Consortium for Top-Down Proteomics

CV: Controlled Vocabulary

FAIR: Findable, Accessible, Interoperable and Reusable

GNO: Glycan Naming Ontology

HUPO: Human Proteome Organization

IUPAC: International Union of Pure and Applied Chemistry

MS: Mass Spectrometry

PDB: Protein Data Bank

PEFF: PSI Extended FASTA Format

PFR: Proteoform Record

PSI: Proteomics Standards Initiative

PrSM: Proteoform Spectrum Match

PSM: Peptide-Spectrum Match

PTM: Post-Translational Modification

SDK: Software Development Kit

UniProtKB: UniProt Knowledge-Base

USI: Universal Spectrum Identifier

**Supplemental Information for: "Proteomics Standards Initiative's ProForma 2.0:**

**Unifying the Encoding of Proteoforms and Peptidoforms"**

Richard D. LeDuc[1,2*], Eric W. Deutsch[3], Pierre-Alain Binz[4], Ryan T. Fellers[1], Anthony J. Cesnik[5,6,7], Joshua A. Klein[8], Tim Van Den Bossche[9,10], Ralf Gabriels[9,10], Arshika Yalavarthi[1], Yasset Perez-Riverol[11], Jeremy Carver[12,13,14], Wout Bittremieux[12,14], Shin Kawano[15,16], Benjamin Pullman[12,13], Nuno Bandeira[12,13,14], Neil L. Kelleher[1], Paul M. Thomas[1,17], Juan Antonio Vizcaíno[11,*]

[1] National Resource for Translational and Developmental Proteomics, Northwestern University, Evanston, IL, 60611, USA.

[2] Present Address: Children's Hospital Research Institute of Manitoba, John Buhler Research Centre, 715 McDermot Avenue, Winnipeg, Manitoba, R3E 3P4.

[3] Institute for Systems Biology, Seattle WA 98109, USA.

[4] Clinical Chemistry Service, Lausanne University Hospital, 1011 Lausanne, Switzerland, 1007.

[5] Department of Genetics, Stanford University, Stanford, CA 94305, USA.

[6] Chan Zuckerberg Biohub, 499 Illinois St, San Francisco, CA 94158, USA.

[7] SciLifeLab, School of Engineering Sciences in Chemistry Biotechnology and Health, KTH – Royal Institute of Technology, SE-171 21 Solna, Stockholm, Sweden, 113 51.

[8] Program for Bioinformatics, Boston University, Boston, MA 02215, USA.

[9] VIB – UGent Center for Medical Biotechnology, VIB, Technologiepark 75 - FSVM II, 9052 Ghent, Belgium.

[10] Department of Biomolecular Medicine, Faculty of Medicine and Health Sciences, Ghent University, 9000 Ghent, Belgium.

[11] European Molecular Biology Laboratory, EMBL-European Bioinformatics Institute (EMBL-EBI), Hinxton, Cambridge, CB10 1SD, United Kingdom.

[12] Center for Computational Mass Spectrometry, University of California, San Diego (UCSD), La Jolla, CA 92093, USA.

[13] Dept. Computer Science and Engineering, University of California, San Diego (UCSD), La Jolla, CA 92093, USA.

[14] Skaggs School of Pharmacy and Pharmaceutical Sciences, University of California, San Diego (UCSD), La Jolla, CA 92093, USA.

[15] Toyama University of International Studies, Toyama. 930-1292 Toyama, Higashikuromaki, 6 5-1, Japan

[16] Database Center for Life Science, Joint Support-Center for Data Science Research, Research Organization of Information and Systems, Kashiwa, Chiba 277-0871, Japan.

[17] Present Address: AbbVie, Inc., 1401 Sheridan Rd, North Chicago, IL 60064 USA.

[*] Corresponding authors: Richard D. LeDuc (RLeduc@chrim.ca) & Juan Antonio Vizcaíno (juan@ebi.ac.uk)

**Supplementary Document 1:** ProForma 2.0 specification document.

Juan Antonio Vizcaíno, EMBL-EBI
Eric W. Deutsch, Institute for Systems Biology
Pierre-Alain Binz, CHUV Lausanne University Hospital
Ryan T. Fellers, Northwestern University
Anthony J. Cesnik, Stanford University
Joshua A. Klein, Boston University
Tim Van Den Bossche, Ghent University
Ralf Gabriels, Ghent University
Yasset Perez-Riverol, EMBL-EBI
Jeremy Carver, University of California San Diego
Shin Kawano, Toyama University of International Studies
Benjamin Pullman, University of California San Diego
Nuno Bandeira, University of California San Diego
Paul M. Thomas, Northwestern University
Richard D. LeDuc, Northwestern University

February 3, 2022

## ProForma 2.0 (Proteoform and Peptidoform Notation)

Status of this document

This document provides information to the proteomics community about a proposed extension of the standard proteoform notation called ProForma. Distribution is unlimited.

Version Draft 15 - this is a draft of version 2.0.0

Abstract

The Human Proteome Organisation (HUPO) Proteomics Standards Initiative (PSI) defines community standards for data representation in proteomics to facilitate data comparison, exchange and verification. This document presents a specification for a proteoform and peptidoform notation, which is based on the ProForma notation [1], previously published by the Consortium for Top-Down Proteomics.

Further detailed information, including any updates to this document, implementations, and examples is available at http://psidev.info/proforma.

Contents

## 1. Introduction

### 1.1 Description of the need

Protein and peptide sequences are usually represented using a string of amino acids using a well-known one letter code endorsed by the IUPAC (see e.g. https://wissen.science-and-fun.de/chemistry/biochemistry/iupac-one-letter-codes-for-bioinformatics/). Representing all the possible variations of a protein or peptide primary structure, including both artefactual and post-translational modifications (PTMs) of peptides and proteins is less clear. For example, the Consortium for Top-Down Proteomics (CTDP) has introduced a standard proteoform notation format called ProForma [1, 2] for writing the primary structures of fully characterized proteoforms [3]. Proteoforms comprise protein species that include variations arising from genetic, transcriptomic, translational, post-translational, and artefactual (e.g., during sample processing) sources. ProForma specifically focuses on representing post-translational modifications of endogenous and artefactual sources. Briefly, ProForma describes proteoforms as the amino acid sequences (the one-letter code representation) complemented with information on any modifications (of a known identity or via unidentified mass shifts) given in brackets following certain amino acids.

Despite its suitability to support a wide range of possible use cases, the original ProForma notation had some limitations. Additionally, the Proteomics Standards Initiative (PSI) has developed a format called PEFF (PSI Extended FASTA Format, http://www.psidev.info/peff) [4]. Although PEFF's primary intended use is for representing search databases for optimising proteomics analyses, PEFF can also be used to represent proteoforms [3] (see more details in Section 4). Therefore, there are multiple ways of encoding protein modifications and extended discussion has taken place to achieve a consensus. A comprehensive standard notation for proteoforms, as well as for their peptidic counterparts –peptidoforms (term introduced in [5])– is then required for the community, so that it can enhance the current description or be newly embedded in many relevant PSI (and potentially other) file formats.

The format specification presented here, ProForma 2.0, represents the consensus between both groups, CTDP and PSI, for the enhanced standard representation of proteoforms and peptidoforms. Compared to the original ProForma notation, it aims to support a broader variety of peptidomics and proteomics approaches, including bottom-up (focused on peptides/peptidoforms) and middle/top-down (focused on proteins/proteoforms) approaches [6]. The name of the notation, ProForma 2.0, derives from the original ProForma notation introduced by CTDP. For simplicity, going forward we will refer to this extended notation as ProForma.

### 1.2 Requirements

The main eight requirements to be fulfilled for a proteoform and peptidoform notation are:

- It MUST be a string that is human readable, so it can be generally understood by human individuals.

- It MUST be machine parsable. Other variants of this notation will not be supported computationally, although they could be 'human readable.'
- It MUST be able to support the encoding of amino acid sequences and protein modifications.
- It MUST be able to support the main use cases needed by the proteomics community as a whole, including both bottom-up (focused on peptides/peptidoforms) and middle/top-down (focused on proteins/proteoforms) approaches.
- It MUST be flexible to accommodate different "flavours" of notations, considering common current use.
- It MUST be compatible with existing PSI file formats, where it could be used.
- It MUST be able to capture ambiguity in the position of the modified sites.
- It MUST be able to evolve, so new use cases can be added iteratively in the future.

Several of these requirements, particularly the first three, coincide with those of the original ProForma notation [2]. The fourth requirement was present in the ProForma notation description, but now includes support for the bottom-up proteomics-specific entities, i.e., peptides, whereas the original ProForma notation exclusively targeted whole proteoforms. The final four requirements are new.

## 1.3    Issues to be addressed

The main issues to be addressed by ProForma are:

- It MUST be able to represent peptidoforms and proteoforms in a consistent and reproducible way, considering the different ways of representing protein modifications.
- It MUST be able to be used jointly with the Universal Spectrum Identifier (USI), to represent peptide-spectrum matches (PSMs), and to represent proteoform-spectrum matches (PrSMs).

## 2.    Notational Conventions

The key words "MUST", "MUST NOT", "REQUIRED", "SHALL", "SHALL NOT", "SHOULD", "SHOULD NOT", "RECOMMENDED", "MAY", and "OPTIONAL" are to be interpreted as described in RFC 2119 (2).

## 3.    The Proteoform and Peptidoform Notation Definition

## 3.1    The documentation

The documentation of the ProForma Notation for proteoform and peptidoforms is divided into several components. All components in their most recent form are available at the HUPO-PSI website (http://psidev.info/proforma) and at the ProForma GitHub page (https://github.com/HUPO-PSI/ProForma/).

- Main specification document (this document).

- List of current implementations with examples.
    - C# ProForma Parser: https://github.com/topdownproteomics/sdk
    - USI implementation (Institute for Systems Biology, http://proteomecentral.proteomexchange.org/usi/).

## 3.2    Relationship to other specifications

The format specification described in this document is not being developed in isolation; indeed, it is designed to be complementary to, and thus used in conjunction with, several existing and emerging models. Related specifications include the following:

1. *PSI Universal Spectrum Identifier* (http://www.psidev.info/USI). The PSI Universal Spectrum Identifier is designed to provide a universal mechanism for referring to a specific spectrum in public repositories. It can optionally include an interpretation of the spectrum using the notation described in this specification. Displayers of USIs MAY use any of the supported ProForma notations.

2. *mzSpecLib, the PSI spectrum library format* (http://psidev.info/mzSpecLib). The PSI spectrum library format is being developed as a standard mechanism for storing spectrum libraries. Identified spectra of modified peptides, will have to include the modification information, potentially in this ProForma notation. Furthermore, many spectrum library entries are derived from multiple spectra, and this provenance will be referenced using USIs.

3. *PROXI* (http://www.psidev.info/proxi). The Proteomics Expression Interface being developed by the PSI is a standardized API by which mass spectrometry proteomics information can be exchanged. References to individual spectra will be made via USIs.

4. *PEFF* (http://www.psidev.info/peff). Although it is not its main intended use, the PSI Extended Fasta Format enables the representation of proteoforms [4]. However, PEFF was not designed for the representation of the (potentially much shorter) peptidoforms.  Additionally, PEFF 1.0 supports formally only a subset of the use cases outlined in this specification. Another key difference is that each proteoform instance in PEFF requires a FASTA header, whereas this is not required in ProForma.

5. *ProForma* (http://psidev.info/proforma). ProForma Proteoform Notation version 1, which enables the representation of proteoforms (https://topdownproteomics.github.io/ProteoformNomenclatureStandard/), developed by the CTDP [1]. This specification is subsumed by this new version 2 ProForma specification.

## 4.   The Basic Form of the Proteoform and Peptidoform Notation

The ProForma notation is a string of characters that represent linearly one or more peptidoform/proteoform primary structures with possibilities to link peptidic chains together. It is not meant to represent higher order structures.

ProForma is case insensitive. However, within the data that follows the different keys, capitalisation may be important. In that case, capitalisation sensitivity is the decision of the supported CVs/ontologies.

Since ProForma MAY be used to represent both peptidoforms and proteoforms, there is currently no limit in its maximum length. Line breaks MUST NOT be used. However, non-ASCII characters are also allowed since non-ASCII characters can be included in the supported ontologies and controlled vocabularies (CVs).

If implementers want to add any metadata (e.g. date of creation, software, version of ontologies, etc) to ProForma entities, the way to do it in this version would be to use the INFO tag.

Due to the multiple use cases supported in this specification, it is not expected that all implementers can provide support to all the supported features from ProForma. To facilitate adoption and separate some of the use cases, there are multiple "levels of compliance" and extensions for ProForma, which are summarised in Appendix I.

## 4.1   The canonical amino acid sequence

Amino acid sequences are represented by strings of amino acids represented as characters using the one letter code endorsed by the IUPAC (http://publications.iupac.org/pac/1984/pdf/5605x0595.pdf
 and https://wissen.science-and-fun.de/chemistry/biochemistry/iupac-one-letter-codes-for-bioinformatics/). There are also letters for representing ambiguous and/or unusual amino acids (see http://www.insdc.org/documents/feature_table.html#7.5.3), which are used in some UniProt entries. Some examples are:

- B: Aspartic Acid or Asparagine
- Z: Glutamic Acid or Glutamine
- J: Leucine or Isoleucine
- U: Selenocysteine
- O: Pyrrolysine
- X: Any amino acid (see also Section 4.2.6 Specifying a gap of known mass, for the use of X). We note that the character X itself is assigned zero mass in this notation.

The representation of non-linear peptides is NOT formalised in this version of ProForma. See the section 5.3 ("Representation of cyclic peptides") in *Section 5: Pending Issues*, for possible ways to represent them.

## 4.2    Generic representation of protein modifications

It has been decided that multiple formats and reference systems must be supported, because some flexibility is required. The same approach is followed for both artefactual protein modifications and natural PTMs. Square brackets MUST be used to represent them when the position is unambiguous. They are located after the character representing the modified amino acid. If there is ambiguity in the position of the protein modification, different rules apply (see section 3.3.4).

Five different reference systems for protein modifications are supported including the following CVs and/or ontologies:

- Unimod (http://www.unimod.org/).
- PSI-MOD (https://github.com/HUPO-PSI/psi-mod-CV).
- RESID (https://proteininformationresource.org/resid/). Although RESID is included in PSI- MOD, this reference system is still used in the top-down community.
- XL-MOD (https://raw.githubusercontent.com/HUPO-PSI/mzIdentML/master/cv/XLMOD.obo) MUST be used for the representation of cross-linkers.
- GNO (Glycan Naming Ontology, https://www.ebi.ac.uk/ols/ontologies/gno).

### 4.2.1    Controlled vocabulary or ontology modification names

The names from different CV or ontology terms MAY be used to represent protein modifications. The two main reference systems used are Unimod and PSI-MOD. However, to facilitate differentiation between reference systems for readers, the names coming from other three supported CV/ontology MUST be preceded by a letter and colon, indicating the originating CV/ontology. In the case of Unimod and PSI-MOD, the use of prefixes is optional.

Examples of proper modification name usage:
- Unimod: U (optional)
- PSI-MOD: M (optional)
- RESID: R (mandatory)
- XL-MOD: X (mandatory)
- GNO: G (mandatory)

EM[Oxidation]EVEES[Phospho]PEK (example using Unimod names)
EM[L-methionine sulfoxide]EVEES[O-phospho-L-serine]PEK (example using PSI-MOD names)
EM[R: L-methionine sulfone]EVEES[O-phospho-L-serine]PEK
EMEVTK[X:DSS#XL1]SESPEK (see Section 4.2.3)

In the case of GNO, the use of accession numbers is preferred since accession numbers and names are often the same. Example:

NEEYN[GNO:G59626AS]K is preferred over NEEYN[G:G59626AS]K

Prefixes can still be used for Unimod and PSI-MOD names (but it is not included in basic support, see Appendix I):

EM[U:Oxidation]EVEES[U:Phospho]PEK
EM[M:L-methionine sulfoxide]EVEES[M:O-phospho-L-serine]PEK

If prefixes are not used for CV/ontology term names, different CVs/ontologies in the same ProForma instance SHOULD NOT be mixed:

EM[U:Oxidation]EVEES[M:O-phospho-L-serine]PEK
EM[Oxidation]EVEES[O-phospho-L-serine]PEK -> Different CVs/ontologies SHOULD NOT be used.

Special characters do not need to be escaped. The only restriction is that unpaired bracket characters MUST NOT be used. Example of properly paired internal brackets:

EM[Oxidation]EVE[Cation:Mg[II]]ES[Phospho]PEK

For different reference systems not supported explicitly, the tag 'INFO' MUST be used (see Section 4.7).

### 4.2.1.1   Definition of the Unimod modification name

The Unimod OBO file SHOULD be used: http://www.unimod.org/obo/unimod.obo. Within this file, term names are found in the "name" tag. These terms differ in the Unimod web interface (http://www.unimod.org/). There, the equivalent to the "name" field in the OBO file is the "PSI-MS Name" column, if not empty (if there is a value). If the "PSI-MS Name" field is empty, the "interim name" is used. Unimod synonyms are currently NOT supported, as they are provided inconsistently.

### 4.2.2   Controlled vocabulary or ontology protein modification accession numbers

In case accession numbers from the supported CVs/ontologies are used, to report protein modifications full accession numbers MUST be used in all cases (no abbreviations in the names of the ontologies/CVs are allowed). The supported names are:
- Unimod: UNIMOD
- PSI-MOD: MOD
- RESID: RESID
- XL-MOD: XLMOD
- GNO: GNO

Examples of proper accession number usage:

EM[MOD:00719]EVEES[MOD:00046]PEK
EM[UNIMOD:35]EVEES[UNIMOD:56]PEK
EM[RESID:AA0581]EVEES[RESID:AA0037]PEK

The following examples are incorrect:

EM[M:00719]EVEES[M:00046]PEK
EM[U:35]EVEES[U:56]PEK
EM[R:AA0581]EVEES[R:AA0037]PEK

### 4.2.3   Support for cross-linkers

Support for cross-linkers is possible by using the XL-MOD CV. It is acknowledged that the current version of ProForma does not provide support for all possible use cases involving cross-linked peptides. In the future, it is expected that a specific extension for this type of information can be developed.

Using the XL-MOD CV, crosslinked sites MUST represented immediately following the modification notation using the prefix #XL, followed by an arbitrary label consisting of alphanumeric characters ([A-Za-z0-9]+ in regular expression notation). Cross-linker modification notations MUST be mentioned once only.

Any annotation made with the symbol # represents a way of linking different locations within the amino acid sequence. In ProForma 2.0 it is used for representing cross-linkers, branched peptides and for grouping protein modifications (including glycans) to represent ambiguity.

### 4.2.3.1   Crosslink notation (within the same peptide)

Cross-linker modification notations MUST be mentioned once only. This example shows a DSS crosslink between two lysines:

EMEVTK[XLMOD:02001#XL1]SESPEK[#XL1]

This second example shows a DSS crosslink between two lysines and an EDC cross-link between two other lysines:

EMK[XLMOD:02000#XL1]EVTKSE[XLMOD:02010#XL2]SK[#XL1]PEK[#XL2]AR

A "dead end" crosslink happens regularly with bifunctional crosslinkers when one side attaches and the other hydrolyses before attaching. These modifications are annotated at only one site.

EMEVTK[XLMOD:02001#XL1]SESPEK

EMEVTK[XLMOD:02001]SESPEK

**4.2.3.2   Representing inter-chain crosslinks**

Inter-protein or inter-chain connections are supported using // to separate the crosslinked peptides. This notation is similar to IUPAC condensed notation for inter-protein connections.

SEK[XLMOD:02001#XL1]UENCE//EMEVTK[XLMOD:02001#XL1]SESPEK

SEK[XLMOD:02001#XL1]UENCE//EMEVTK[#XL1]SESPEK

It is acknowledged by the authors that more complex scenarios are possible when representing inter-chain crosslinks, including a higher number of linked peptides, directionality, etc. It is envisioned that when these use cases become a clear requirement in the future, a dedicated working group can extend these guidelines.

**4.2.3.3   Representing disulfide linkages**

Disulfide bonds may be represented using four possible notations:

(i) Using the PSI-MOD term for "L-cystine (cross link)" (MOD:00034) to explicitly describe the cross-link using the cross-linking notation:

EVTSEKC[MOD:00034#XL1]LEMSC[#XL1]EFD
EVTSEKC[L-cystine (cross-link)#XL1]LEMSC[#XL1]EFD

There are more complex examples that are possible. For instance, another example with inter-chain disulfide bonds is insulin:

FVNQHLC[MOD:00034#XL1]GSHLVEALYLVC[MOD:00034#XL2]GERGFFYTPK
A\\GIVEQC[MOD:00034#XL3]C[#XL1]TSIC[#XL3]SLYQLENYC[#XL2]N

As mentioned above, more complex scenarios are possible which will need to be resolved in future versions.

(ii) Using the XLMOD term XLMOD:02009 similarly to case (i) above:

EVTSEKC[XLMOD:02009#XL1]LEMSC[#XL1]EFD
EVTSEKC[X:Disulfide#XL1]LEMSC[#XL1]EFD

(iii) Using the PSI-MOD term for "half cystine" (MOD:00798) if the pairing is not known. Since the term is only for half the link, it must be specified on all involved sites with no group tag:

EVTSEKC[half cystine]LEMSC[half cystine]EFD
EVTSEKC[MOD:00798]LEMSC[MOD:00798]EFDEVTSEKC[MOD:00798]LEMSC[MOD:00798]EFD

(iv) Using the Unimod term for "Dehydro" (UNIMOD:374) to explicitly describe the cross-link using the cross-linking notation.

EVTSEKC[UNIMOD:374#XL1]LEMSC[#XL1]EFD
EVTSEKC[Dehydro#XL1]LEMSC[#XL1]EFD

### 4.2.4    Representation of branched peptides

Branched peptides can be expressed using the same notation used for representing two cross-linked peptides, but using the term #BRANCH (see above). Examples:

a) ETFGD[MOD:00093#BRANCH]\\R[#BRANCH]ATER

b) Cross-linked via a sidechain:

AVTKYTSSK[MOD:00134#BRANCH]\\AGKQLEDGRTLSDYNIQKESTLHLVLRLR
G-[#BRANCH]
Where a sidechain of a Lysine from peptide 1 is linked to the C-term of the peptide 2 via amidation (-H2O)

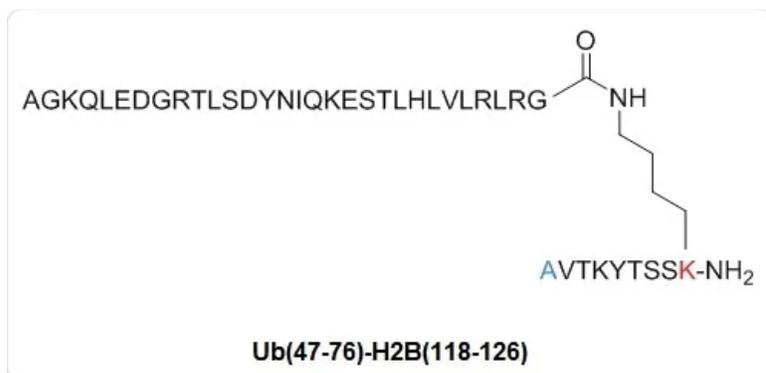

**Ub(47-76)-H2B(118-126)**

Taken from:
https://www.news-medical.net/whitepaper/20180329/Synthesizing-Unsymmetrically-Branched-Peptides.aspx

### 4.2.5    Representation of glycans using the GNO ontology as CV

Glycans that are currently included in Unimod or PSI-MOD (individual or very short chains) MAY be represented that way. If the glycans are not included in either PSI-MOD or Unimod, the GNO ontology SHOULD be used. As mentioned above, the use of accession numbers is preferred since accession numbers and names are often the same.

Examples of proper glycan notation:

Encoding  Hex 5 HexNAc 4 NeuAc 1:
NEEYN[GNO:G59626AS]K

http://psidev.info/proforma

Encoding Hex 8 HexNAc 2 and Hex 5 HexNAc 2:
YPVLN[GNO:G62765YT]VTMPN[GNO:G02815KT]NSNGKFDK

The same mechanisms for expressing labile modifications and ambiguity in the modification position applicable to other types of modifications SHOULD be used for glycans as well (see following sections, e.g. Sections 4.3.2 and 4.4).

There are more complex cases, where ambiguity can be caused by multiple combinations between labile and non-labile glycans attached to the same amino acid sequence. A possible mechanism to represent these more complex cases is available in Section 5 (*Pending issues*). A further limitation comes from the restricted set of glycans in GNO. We expect that these issues will be solved as the glyco(proteomics) community develops in the near future.

### 4.2.6   Delta mass notation

In addition to using CV/ontologies names and/or accession numbers, mass differences (delta masses) MAY be used to represent protein modifications.

Delta masses SHOULD only be used when the protein modification cannot be represented using a CV/ontology (e.g., if software does not use ontologies/CVs), when the modification (or combination of modifications) is ambiguous (e.g., coming from open modification searches or de-novo approaches), or when it is unknown. Otherwise, protein modifications SHOULD be represented using Unimod, PSI-MOD, RESID, XL-MOD, or GNO CV parameters.

Mass differences MUST be expressed in Daltons between the coded amino acid and the observed mass. Positive mass shifts MUST be specified with a plus sign. Negative shifts MUST be specified with a negative sign. Monoisotopic masses MUST be used. There are two ways of representing delta masses:

A) Without using prefixes.

EM[+15.9949]EVEES[+79.9663]PEK
EM[+15.995]EVEES[-18.01]PEK

Interpretation of the actual delta masses is then left to the reader software.

B) Using prefixes for CVs/ontologies to provide more information.

If "canonical" delta masses are directly taken from a CV/ontology, the corresponding abbreviation to that CV/ontology MAY be used.

- Unimod: U
- PSI-MOD: M

- RESID: R
- XL-MOD: X
- GNO: G

Examples of delta masses corresponding to CV/ontology entries:

EM[U:+15.9949]EVEES[U:+79.9663]PEK
EM[U:+15.995]EVEES[U:+79.966]PEK

The notation also supports the encoding of experimentally observed delta masses. In those cases, the prefix "Obs" MUST be used. The number of significant figures included in the delta mass depends on the accuracy of the available data and SHOULD be used as is by interpreters. Example:

EM[U:+15.995]EVEES[Obs:+79.978]PEK

### 4.2.7 Specifying a gap of known mass

This mechanism can be used to express a gap in the sequence of an unknown number of amino acids, but the corresponding mass difference is known. This is supported by the use of the character X followed by brackets indicating the total mass of the gap, meaning that the mass of X is actually zero.

Example of proper gap notation:

RTAAX[+367.0537]WT

### 4.2.8 Support for elemental formulas (e.g. for representing small molecular substructures or functional groups)

A modification representing a small molecular substructure or a functional group can be described by a chemical formula. The descriptor "Formula" MUST be used. Only elemental formulas are supported. Example of proper chemical formula usage:

SEQUEN[Formula:C12H20O2]CE
SEQUEN[Formula:[13C2]CH6N]CE

As no widely accepted specification exists for expressing elemental formulas, we have adapted a standard with the following rules (taken from https://github.com/rfellers/chemForma):

Formula Rule 1
A formula will be composed of pairs of atoms and their corresponding cardinality (two Carbon atoms: C2). Pairs SHOULD be separated by spaces but are not required to be. Atoms and cardinality SHOULD NOT be. Also, the Hill system for ordering

([https://en.wikipedia.org/wiki/Chemical_formula#Hill_system](https://en.wikipedia.org/wiki/Chemical_formula#Hill_system)) is preferred, but not required.

Example: C12H20O2    or    C12 H20 O2

Formula Rule 2
Cardinalities must be positive or negative integer values. Zero is not supported. If a cardinality is not included with an atom, it is assumed to be +1.

Example: HN-1O2

Formula Rule 3
Isotopes will be handled by prefixing the atom with its isotopic number in square brackets. If no isotopes are specified, previous rules apply. If no isotope is specified, then it is assumed the natural isotopic distribution for a given element applies.

Example: [13C2][12C-2]H2N
Example: [13C2]C-2H2N

SEQUEN[Formula:[13C2][12C-2]H2N]CE
(here 2 $^{12}C$ atoms are replaced by 2 $^{13}C$ atoms)

See in Section 5 (*Pending issues*) how this mechanism could be extended in the future to support more complex molecular formulas.

### 4.2.9   Representation of glycan composition

Glycan residues (generic monosaccharides) can be represented using the descriptor "Glycan". If glycan symbols conflict with themselves or element symbols in such a way that ambiguities occur, we will consider requiring spaces between 'atoms' (see Formula Rule #1).

Example: Hex2HexNAc

SEQUEN[Glycan:HexNAc1Hex2]CE

The supported list of monosaccharides in ProForma is included below. It is worth noting that the masses and elemental compositions included below for each monosaccharide are those resulting after each of them are condensed with the amino acid chain.

Hex: Hexose, 162.0528 Da, C6H10O5
HexNAc: N-Acetyl Hexose, 203.0793 Da, C8H13N1O5
HexS: Hexose Sulfate, 242.0096 Da, C6H10O8S1
HexP: Hexose Phosphate, 242.0191 Da, C6H11O8P1
HexNAcS: N-Acetyl Hexose Sulfate, 283.0361 Da, C8H13N1O8S1
dHex: Deoxy-Hexose, 146.0579 Da, C6H10O4

NeuAc: N-acetyl Neuraminic Acid / Sialic Acid, 291.0954 Da, C11H17N1O8
NeuGc: N-glycolyl Neuraminic Acid, 307.0903 Da, C11H17N1O9
Pen: Pentose, 132.0422 Da, C5H8O4
Fuc: Fucose, 146.0579 Da, C6H10O4 (a particular stereochemical assignment of dHex abundant in mammals)

However, we envision that more monosaccharides could be added once this specification document is formalised. An updated list of supported monosaccharides (in two different formats, obo and json) can be found at:

https://github.com/HUPO-PSI/ProForma/tree/master/monosaccharides

For other glycans not included there, a new CV term will need to be created, e.g. in PSI-MOD.

It is recognised that this mechanism is limited and can only support the most common glycans. It is envisioned that in the future, when this use case becomes a requirement, a dedicated working group can work in extending these specific guidelines. See Section 5 (*Pending issues*) for guidance on future extensions of this mechanism to support other macromolecules, e.g. lipids.

### 4.2.10  Best practices on the use of protein modifications

In the same sequence, the same reference system SHOULD be used to represent the protein modifications. However, the delta mass notation (Section 4.2.5) MAY be combined with the other cases.

### 4.3     Representation of special cases: N-terminal, C-terminal and labile protein modifications

### 4.3.1     N-terminal and C-terminal modifications

The square brackets containing the modification MUST be located before the first amino acid in the sequence or after the last amino acid in the peptide sequence. In both cases, they are separated by a dash (-). Examples:

[iTRAQ4plex]-EM[Oxidation]EVNES[Phospho]PEK

[iTRAQ4plex]-EM[U:Oxidation]EVNES[Phospho]PEK[iTRAQ4plex]-[Methyl]

### 4.3.2   Labile modifications

Labile modifications are those which are known to separate under certain experimental conditions during fragmentation and therefore are not visible in the fragmentation MS2 spectrum (i.e. the MS2 spectra are indistinguishable from spectrum not containing the modification). They are represented by curly brackets {}, not by square ones. As explained in Section 4.2.8, the prefix "Glycan:" needs to be added for each labile monosaccharide. Labile modification MUST be located before the first amino acid sequence and before N-terminal modifications, if applicable. Examples:

{Glycan:Hex}EM[U:Oxidation]EVNES[Phospho]PEK[iTRAQ4plex]

{Glycan:Hex}[iTRAQ4plex]-EM[Oxidation]EVNES[Phospho]PEK[iTRAQ4plex]

{Glycan:Hex}[iTRAQ4plex]-EM[Oxidation]EVNES[Phospho]PEK[iTRAQ4plex]-[Methyl]

One can also express multiple labile modifications using the following notation:

{Glycan:Hex}{Glycan:NeuAc}EMEVNESPEK

## 4.4    Support for the representation of ambiguity in the modification position

This notation is used to represent ambiguous modified sites, associated positions and associated probabilities or scores.

This notation is not yet supported for crosslinker modifications (see Section 5.9), except for the case of disulfide cross-linkers which may be represented with ambiguous position using the PSI-MOD term for "half-cystine" (MOD:00798), as noted in Section 4.2.3.3.iii.

### 4.4.1    Unknown modification position

The positions of some modifications may be unknown. In this case, protein modifications are represented using square brackets that MUST be located on the left side of the amino acid sequence. The symbol '?' is used to indicate that the actual position of the modification is unknown.

[Phospho]?EM[Oxidation]EVTSESPEK

In case of multiple modifications with an unknown location, two options are possible to represent them:

(i) Listing them separately as in this example of two phosphorylations:

[Phospho][Phospho]?[Acetyl]-EM[Oxidation]EVTSESPEK

(ii) Indicating the concrete modification only once but using the caret (^) symbol to represent the number of occurrences of the modification.

[Phospho]^2?[Acetyl]-EM[Oxidation]EVTSESPEK

N-terminal modifications MUST be the last ones written, just next to the amino acid sequence. For example:

Wrong: [Acetyl]-[Phospho]^2?EM[Oxidation]EVTSESPEK
Right: [Phospho]^2?[Acetyl]-EM[Oxidation]EVTSESPEK

### 4.4.2    Indicating a possible set of modification positions

The position of a modification may be unknown but belong to a known set of possible sites. In this case, the possible positions for the modifications may be indicated. The rules that MUST be followed are:

(i) Groups of possible sites for a modification are represented immediately following the modification notation using the symbol #, followed by an arbitrary label consisting of alphanumeric characters ([A-Za-z0-9]+ in regular expression notation). Note that the label prefix #XL is a special case that MUST be reserved for crosslinkers only.

(ii) A single preferred location for the modification MUST be specified, so that the sequence can be easily rendered in visualization tools. The preferred location for the modification is indicated by the position of the modification notation in the amino acid sequence.

In this example, '#g1' is used as the arbitrary label:

EM[Oxidation]EVT[#g1]S[#g1]ES[Phospho#g1]PEK

This is read as a named group 'g1' indicates that a phosphorylation exists on either T5, S6 or S8, and S8 is the preferred location because the notation 'Phospho' is placed at this position.

The following example is not valid because a single preferred location must be chosen for a modification:

EM[Oxidation]EVT[#g1]S[Phospho#g1]ES[Phospho#g1]PEK

### 4.4.3    Representing ranges of positions for the modifications

Ranges of amino acids as possible locations for the modifications may be represented using parentheses within the amino acid sequence. Some examples:

PRT(ESFRMS)[+19.0523]ISK
PRT(EC[Carbamidomethyl]FRMS)[+19.0523]ISK

The caret symbol (^), which can be used to represent multiple instances of the same unlocalised modification before the N-terminal end of the amino acid sequence (Section 4.4.1), is not allowed within the amino acid sequence.

Overlapping ranges represent a more complex case and are not yet supported, and so, the following example would NOT be valid:

P(RT(ESFRMS)[+19.0523]IS)[+19.0523]K

### 4.4.4    Indicating modification position preference and localisation scores

There are two options to represent this type of information. The values of the modification localisation scores can be indicated in parentheses within the same group and brackets.

Example of proper localisation score usage:

EM[Oxidation]EVT[#g1(0.01)]S[#g1(0.09)]ES[Phospho#g1(0.90)]PEK

Scores for the modification position can be expressed as probabilities and/or FLR (False Localisation Rate), but the actual meaning of the scores is not reported. The preferred location of the modification notation reflects the value of the scores. If there is a tie in the value of the localisation scores, one preferred position needs to be chosen by the writer.

An additional option to represent localisation scores is to leave the position of the modification as unknown using the '?' notation but report the localisation modification scores at specific sites.

Example of proper usage of localisation scores with unknown modification site notation:

[Phospho#s1]?EM[Oxidation]EVT[#s1(0.01)]S[#s1(0.09)]ES[#s1(0.90)]PEK

### 4.4.5    Representing scoring for ranges of positions for a modification

Ranges of amino acids as possible locations for the modifications may also be accompanied by scoring using the same notation. Some examples:

PRT(ESFRMS)[+19.0523#g1(0.01)]ISK[#g1(0.99)]
PR[#g1(0.91)]T(EC[Carbamidomethyl]FRMS)[+19.05233#g1(0.09)]ISK

### 4.5    Representation of multiple modifications in the same amino acid residue

It is possible to represent two or more modifications on the same amino acid or group of amino acids. The caret symbol (^), which can be used to represent multiple instances of the same unlocalised modification before the N-terminal end of the amino acid sequence (Section 4.4.1.) is however not allowed within the amino acid sequence. No extra character is required. Example:

MPGLVDSNPAPPESQEKKPLK(PCCACPETKKARDACIIEKGEEHCGHLIEAHKEC
MRALGFKI)[Oxidation][Oxidation][half cystine][half cystine]

This would not be allowed:

MPGLVDSNPAPPESQEKKPLK(PCCACPETKKARDACIIEKGEEHCGHLIEAHKEC
MRALGFKI)[Oxidation]^2[half cystine][half cystine]

Currently, complex glycans are not explicitly supported (see Section 3.4). An alternative solution in those rare cases not involving glycans is to have a single PSI-MOD/Unimod entry for the combination of mods, which would need to be created in advance, if not yet available.

## 4.6    Representation of global modifications

This mechanism MAY be used for modifications that apply to all relevant residues in the peptide/protein amino acid sequence. These modifications MAY be represented by the use of the characters "<" and ">" on the left side of the sequences. A couple of use cases are envisioned:

### 4.6.1    Use Case 1: Representation of isotopes

This might be used in the case of synthetic peptides with 100% incorporation.

Example: Consider extension for 13C on all residues:
Carbon 13: <13C>ATPEILTVNSIGQLK
Nitrogen 15: <15N>ATPEILTVNSIGQLK
Deuterium: <D>ATPEILTVNSIGQLK

The representation of multiple isotopes is also possible. They can be located in any order.

Both Carbon 13 and Nitrogen 15: <13C><15N>ATPEILTVNSIGQLK

Distributions of isotope masses could be supported in future work.

### 4.6.2    Use Case 2: Fixed protein modifications

This mechanism can be useful especially in the case of full proteoforms. The affected amino acid MUST be indicated using @. If more than one residue were affected, they MUST be comma separated. Examples:

<[S-carboxamidomethyl-L-cysteine]@C>ATPEILTCNSIGCLK
<[MOD:01090]@C>ATPEILTCNSIGCLK
<[Oxidation]@C,M>MTPEILTCNSIGCLK

Fixed modifications MUST be written prior to ambiguous and labile modifications, and similar to ambiguity notation, N-terminal modifications MUST be the last ones written, just next to the sequence.

The following examples would be valid:

<[MOD:01090]@C>[Phospho]?EM[Oxidation]EVTSECSPEK
<[MOD:01090]@C>[Acetyl]-EM[Oxidation]EVTSECSPEK

## 4.7   Representation of amino acid sequence ambiguity

Ambiguity in the amino acid sequence needs to be represented in some cases, e.g. to represent sequence changes that <u>do not change the mass</u> of the peptidoform/proteoform, but are not known. One concrete example is the need to encode the results of *de novo* sequencing tools. The way to encode this information is to use a parenthesis and a quotation mark including the ambiguous sequence represented in a preferred way. Examples:

(?DQ)NGTWEM[Oxidation]ESNENFEGYM[Oxidation]K
(?N)NGTWEM[Oxidation]ESNENFEGYM[Oxidation]K

In both examples, both ambiguous amino acid sequences are DQ and N, respectively.

## 4.8   The information tag

General information or comments can be encoded using the 'info' tag like:
ELV[INFO:AnyString]IS
ELV[info:AnyString]IS

The information represented in an 'info' tag is considered non-standard (e.g. any text besides unpaired brackets) and does not need to be parsed.

"Info" tags can be split using the pipe character. Example of proper 'info' tag usage:

ELVIS[Phospho|INFO:newly discovered]K
ELVIS[Phospho|INFO:newly discovered|INFO:really awesome]K

The following comment would be invalid because of an unpaired bracket:

ELVIS[Phospho|INFO:newly]discovered]K

As a concrete example of its use, "Info" tags can be used to provide metadata about ProForma entities (e.g. date of creation, version of Unimod used, software used for creating it, and many others):

ELVIS[Phospho|INFO:newly discovered|INFO:Created on 2021-06]K
ELVIS[Phospho|INFO:newly discovered|INFO:Created by software Tool1]K

**4.9   Support for the joint representation of experimental data and its interpretation**

The pipe character "|" is used to represent protein modifications simultaneously with CV/ontology names and/or accession numbers, and delta masses. As explained in Section 4.2.6, Delta mass notation, it is possible to represent both canonical delta masses and experimental observations, allowing the representation of both interpretation (using CV/ontology names/accession numbers) and experimental observations (delta masses).

Examples:

ELVIS[U:Phospho|+79.966331]K
Showing both the interpretation and measured mass:
ELVIS[U:Phospho|Obs:+79.978]K

Other combinations between CV/ontology names, accession numbers, and delta masses using synonyms are allowed, though they MUST be synonymous terms. Some examples:

ELVIS[Phospho|O-phospho-L-serine]K
ELVIS[UNIMOD:21|MOD:00046]K
ELVIS[UNIMOD:21|Phospho]K
ELVIS[Phospho|O-phospho-L-serine|Obs:+79.966]K

Ambiguous cases are also allowed because they can be used to represent "comparable" information.

ELVIS[Obs:+79.966|Phospho|Sulfo]K

Highly different modifications SHOULD NOT be joined as it would be difficult for readers to correctly interpret. It is however acknowledged that readers can choose to implement the parsing in different ways. Some tools may always take CV terms, others could take delta masses, and so on.

## 5.   Pending Issues - Future developments

Additionally, there are several use cases that are NOT currently supported in the current version of the specification. These complications are left open in version 2.0 of the specification and will ideally be addressed in future versions, after the community has gained more experience with the common cases. The objective here is to document those cases appropriately and propose some possible solutions for representing the information in future versions of ProForma.

## 5.1   Representation of cyclic peptides

Cyclic peptides are only currently supported if they can be represented using the supported CVs/ontologies for protein modifications. The following examples represent possible ways to represent cyclic peptides, but these solutions need to be formalised and PSI-MOD modifications created.

1) Cyclic peptide with C- and N-termini bound together at the peptide backbone level
Kalata B1 (PubChemID: 46231131, UniProtKB: P56254)
[*MOD:nnnnnn*#XL1]-RNGLPVCGETCVGGTCNTPGCTCSEPVCT-[#XL1]

where *MOD:nnnnnn* would be a new PSI-MOD term to represent backbone cyclisation involving the amidation between a C-terminal carboxylate and a N-terminal amine, with mass difference of O-1H-2 (-18 Da).

2) Cyclic peptide with C- and N-termini bound together at the peptide backbone level with 3 disulfide bonds
Retrocyclin 1 (PubChem ID 16130540). The exact structure is the following:
https://pubchem.ncbi.nlm.nih.gov/compound/Retrocyclin-1#section=Biologic-Description&fullscreen=true
[*MOD:nnnnnn*#XL1]-
RC[MOD:00798.DS1]IC[MOD:00798.DS2]GRGIC[MOD:00798.DS2]RC[MOD:00798.DS1]IC[MOD:00798.DS3]GRGIC[MOD:00798.DS3]-[#XL1]

where *MOD:nnnnnn* would be a new PSI-MOD term to represent backbone cyclisation involving the amidation between a C-terminal carboxylate and a N-terminal amine, with a mass difference of O-1H-2 (-18 Da).

3) Cyclic peptide with C-terminal COOH condensed to a sidechain NH2
3a) peptide with no other PTM
LEIK[N6-(L-asparagyl)-L-lysine#XL1]KIPHDN[#XL1]

3b) A real case scenario: Topitracin (PubChem ID:6474109)

[[N-[2-[1-amino-2-methylbutyl]-4,5-dihydro-4-thiazolyl]carbonyl]-Leucine]-LE[D-Glutamic acid]IK[M:N6-(L-asparagyl)-L-lysine#XL1]K[M:D-Ornithine]I[M:D-alloisoleucine]P[D-Phenylalanine]HD[M:D-Aspartic acid]N[#XL1]

## 5.2    Representation of ambiguity when different glycans are attached to the same amino acid sequence

Multiply glycosylated peptides, especially under vibrational/collisional dissociation, may fragment in ways that allow sequencing the peptide backbone without completely characterizing the glycan sites. Instead, only the aggregate composition can be determined based on the precursor peptide mass. In such cases, only the glycosylation may be known, by motif for N-glycan or there may be several possible sites. Alternatively, the total number of glycosylation sites may be unknown (O-glycans), with the aggregate glycan composition may be spread across positions in unknown proportions.

There is a need to express that a site is a possible glycosylation site as well as a mechanism to express the total amount of glycan composition shared across these sites. The latter is achieved by using a labile modification to prefix the total composition. There are multiple proposals for expressing putative site assignment:

Proposal 1. Use PSI-MOD glycosylated residue modifications.
{Glycan:Hex 10 HexNAc 4}YPVLN[MOD:00006]VTMPN[MOD:00006]NSNGKFDK

This peptide hosts two N-glycans, where the glycan class is known from the required motifs on the sequence, and that it is multiply glycosylated because no single N-glycan with the aggregate composition is biosynthetically feasible. This proposal denotes the inferred glycosylation sites using the PSI-MOD "N-glycosylated residue" term. This forces the reader to treat this group differently, where the modification is inferred to be the labile glycan modification and that the modification may be split amongst each site, assigning zero or more monosaccharides to each group position.

Pros:
- Conveys extra metadata about the glycan type
- Uses an existing term

Cons:
- Introduces new semantics for a modification that is not explicitly conveyed notationally, namely that this modification is not observable, but just encodes positional information.
- For complex and ambiguous O-glycopeptides, this method would pull double-duty with ambiguity notation.

Proposal 2. Use ambiguity groups.
{Glycan:Hex 10 HexNAc 4}YPVLN[#g1]VTMPN[Glycan#g1]NSNGKFDK

The same case with Proposal 1, but instead of adding extra baggage to an existing term, this proposal uses ambiguity groups to denote possible positions, and mark one group with a new "Glycan" key, which adds the same labile modification inference step.

Pros:

- Uses an ambiguity-specific mechanism to signal ambiguity.

Cons:

- Adds a new component to ambiguity group interpretation that parsers must now be prepared to handle.
- No ability to communicate glycan type at the site level.

A complex O-glycopeptide example
{Glycan:Hex 5 HexNAc
5}PEPSTAT[Glycan#g1]IS[#g1]T[#g1]ICS[#g1]S[#g1]T[#g1]RIKES[#g1]IT[#g1]ES[#g1]

Fragmentation may demonstrate that some S/T residues are not putative sites, while the distribution of glycan composition is still not known amongst the remaining sites. The true solution might be:

PEPSTATISTICS[Glycan:HexNAc Hex]S[Glycan:HexNAc Hex]TRIKES[Glycan:HexNAc Hex]IT[Glycan:HexNAc Hex]ES[Glycan:HexNAc Hex]

Or PEPSTATISTICSS[Glycan:HexNAc 2 Hex 2]TRIKES[Glycan:HexNAc Hex]IT[Glycan:HexNAc Hex]ES[Glycan:HexNAc Hex], or any permutation thereof.

Proposal 3. Use cross-linking-like notation
{Glycan:Hex 10 HexNAc
4.G1}YPVLN[Glycan:#G1]VTMPN[Glycan:#G1]NSNGKFDK

The only differentiating feature of this proposal from Proposal 2 is that it isolates the notational change solely within the Glycan tag handling, which reduces the burden on implementers who do not want to support glycosylation.

## 5.3     Representation of rare amino acids not supported by the one letter code

This use case is currently not supported. These SHOULD be handled through their representations in one of the supported ontologies/CVs.

## 5.4     Representation of average masses

During the development of the format, it was acknowledged that, in the case of top-down proteomics approaches, there could be cases where monoisotopic masses are unknown, and then average masses need to be used. At the moment, monoisotopic masses are the only ones formally allowed, but this MAY have to change in future changes.

## 5.5     Representation of lipids

These SHOULD be handled through their representations in one of the supported ontologies/CVs. However, a similar mechanism to the one described in Section 4.2.8, Representation of glycan composition, could be implemented for lipid molecules.

Examples:

SEQUEN[Lipid:OleicAcid]CE
SEQUEN[Lipid:PalmiticAcid]CE

It is envisioned that when this use case becomes a clear requirement in the future, a dedicated working group can extend these specific guidelines.

## 5.6   Distribution of isotopes in the sequence

The representation of the distributions of isotopes for global modifications (Section 4.6) is not supported in the current version of the specification. A mechanism will need to be envisioned to support this use case in future versions.

## 5.7   Representation of molecular formula

Elemental formulas are supported by the current version of ProForma (Section 4.2.7), and molecular formulas may be supported in the future if it would prove helpful. For example, specifying branching in a PTM structure. A molecular formula may include repeated (condensed) sections using parentheses and an extra cardinality.

Examples:
CH3(CH2)4CH3
SEQUEN[Formula:CH3(CH2)4CH3]CE

## 5.8   Representation of an overlapping range of possible modification positions

Notation of ambiguous localization currently supports non-overlapping ranges. A possible representation of overlapping ranges, that may be considered in the future, uses a grouping tag for both parentheses.

Examples:
PROT([#g1]EOC[Carbamidomethyl]FORMS)[+19.0523#g1]ISK
PR([#g1]OT([#g2]EOC[Carbamidomethyl]FOR)[+19.0523#g1]MS)[+19.0523#g2]ISK
PROT([#g1]([#g2]EOC[Carbamidomethyl]FORMS)[+19.0523#g2]IS)[+19.0523#g1]K

## 5.9   Representation of ambiguous crosslinker modification positions

Notation for ambiguous crosslinker modification positions is not supported in this version of ProForma but may be supported in the future.

## 5.10   Metadata related to ProForma entries

At present, metadata related to ProForma entries (e.g. date of creation, software, version of ontology used, etc) cannot be provided in a standardised manner. The creation of an additional metadata file could be considered in future versions.

### 5.11  Representation of sequences coming from non-mass spectrometry-based proteomics approaches

The Proforma notation could be made compatible with non-mass spectrometry proteomics approaches, such as nanopore and Edman-based sequencing, and other that will face the same notation challenges. A mechanism will need to be envisioned to support these use cases in future versions.

## 6. Appendix I. Levels of Compliance

Due to the multiple use cases supported in this specification, it is not expected that all implementers can provide support to all the supported features from ProForma version 2. To facilitate adoption and separate some of the use cases, there are multiple "levels of compliance" and extensions for ProForma. The technical name of the level of compliance is indicated between parenthesis to enable labelling future software.

1) Base Level Support (Technical name: Base-ProForma Compliant)
Represents the lowest level of compliance, this level involves providing support for:
- Amino acid sequences
- Protein modifications using two of the supported CVs/ontologies: Unimod and PSI-MOD.
- Protein modifications using delta masses (without prefixes)
- N-terminal, C-terminal and labile modifications.
- Ambiguity in the modification position, including support for localisation scores.
- Ambiguity in the amino acid sequence.
- INFO tag.

2) Additional Separate Support (Technical name: level 2-ProForma compliant)
These features are independent from each other:
- Unusual amino acids (O and U).
- Ambiguous amino acids (e.g. X, B, Z). This would include support for sequence tags of known mass (using the character X).
- Protein modifications using delta masses (using prefixes for the different CVs/ontologies).
- Use of prefixes for Unimod (U:) and PSI-MOD (M:) names.
- Support for the joint representation of experimental data and its interpretation.

3) Top-Down Extensions (Technical name: level 2-ProForma + top-down compliant)
- Additional CV/ontologies for protein modifications: RESID (the prefix R MUST be used for RESID CV/ontology term names)
- Chemical formulas (this feature occurs in two places in this list).

4) Cross-Linking Extensions (Technical name: level 2-ProForma + cross-linking compliant)

- Cross-linked peptides (using the XL-MOD CV/ontology, the prefix X MUST be used for XL-MOD CV/ontology term names).

5) Glycan Extensions (Technical name: level 2-ProForma + glycans compliant)
- Additional CV/ontologies for protein modifications: GNO (the prefix G MUST be used for GNO CV/ontology term names)
- Glycan composition.
- Chemical formulas (this feature occurs in two places in this list).

6) Spectral Support (Technical name: level 2-ProForma + mass spectrum compliant)
- Charge and chimeric spectra are special cases (see Appendix II).
- Global modifications (e.g., every C is C13).

If one implementation supports more than one extension, multiple supported extensions can be indicated separated by '+'). Example: (level 2-ProForma + cross-linking + mass spectrum compliant).

Additionally, see Section 5 "Pending Issues - Future developments" for features not yet formally supported in this version of the specification. In the future, there could be additional extensions, e.g., for lipid molecules.

## 7.    Appendix II: Extensions to improve the representation of PSMs in mass spectra

This appendix is not relevant for the representation of peptidoforms and proteoforms, but rather presents techniques for representing PSMs (that is, peptidoforms and proteoforms together with mass spectra).

### 7.1    Representation of the ion charges

The charge value MAY be optionally indicated in the C-terminal end of the amino acid sequence, by using the forward slash (/) character. Examples:

EMEVEESPEK/2
EM[U:Oxidation]EVEES[U:Phospho]PEK/3
[U:iTRAQ4plex]-EM[U:Oxidation]EVNES[U:Phospho]PEK[U:iTRAQ4plex]-[U:Methyl]/3

By default, a positive number n will imply a molecular ion that is n-times protonated SEQUENCE/2 Means [SEQUENCE(neutral)  + 2 protons ] and is doubly charged: $[M+2H^+]^{2+}$

By default, a negative number n will imply a molecular ion that is n-times deprotonated SEQUENCE/-2 Means [SEQUENCE(neutral)  - 2 protons ] and is doubly charged: $[M-2H^+]^{2-}$

When the charge derives from the addition or the removal of another ion, this ionic species SHOULD be provided after the charge state number. Examples include a Na+ adduct, the addition of one electron, the removal of a OH-, the addition of an iodine ion, and a radicalisation.

EMEVEESPEK/2[+2Na+,+H+]
EMEVEESPEK/1[+2Na+,-H+]
EMEVEESPEK/-2[2I-]
EMEVEESPEK/-1[+e-]

### 7.2    Representation of multiple peptidoform assignments in chimeric spectra

In bottom-up approaches, in the case of chimeric spectra, more than one peptidoform sequence MAY be potentially assigned to a single mass spectrum. In this case, multiple peptidoform sequences MUST be separated by the plus sign (+). Example:

EMEVEESPEK/2+ELVISLIVER/3

## 8.  Appendix III. Glossary of terms used in the specification

The objective here is to provide a list of the keys used in the document, so that a summary view is available for implementers.

1- Protein modifications

1.1- (Non-labile) protein modifications are represented by using brackets [] + CV/ontology parameter names (for PSI-MOD/ Unimod).
For RESID (R:), XL-MOD (X:) and GNO (G:), extra prefixes MUST be used before the CV parameter names. For PSI-MOD (M:) and Unimod (U:), they are optional.

EM[Oxidation]EVEES[Phospho]PEK
EM[R: Methionine sulfone]EVEES[O-phospho-L-serine]PEK
EMEVTK[X:DSS#XL1]SESPEK (see Section 4.2.3)
EM[U:Oxidation]EVEES[U:Phospho]PEK

1.2- Non-labile protein modifications can also be reported using brackets [] including +/- Delta mass values.

EM[+15.9949]EVEES[+79.9663]PEK

The use of prefixes for reporting delta masses coming from ontologies/CVs MAY be supported (only in Advanced mode).

EM[U:+15.995]EVEES[U:+79.966]PEK

Experimentally observed delta masses are reported using the prefix [Obs:].

EM[U:+15.995]EVEES[Obs:+79.978]PEK

1.3- Sequence gaps of known mass MAY also be indicated using the amino acid X + brackets [] including the delta mass value of the tag.

RTAAX[+367.0537]WT

1.4- (Labile) protein modifications are indicated at the left side of the sequence using curly brackets {}.

{Glycan:Hex}EM[Oxidation]EVNES[Phospho]PEK[iTRAQ4plex]

1.5- N-terminal and C-terminal modifications are indicated using a dash (-) on the left/right part of the sequence, respectively.

[iTRAQ4plex]-EM[Oxidation]EVNES[Phospho]PEK
[iTRAQ4plex]-EM[Oxidation]EVNES[Phospho]PEK[iTRAQ4plex]-[Methyl]

1.6- Representation of global fixed modifications uses the "at" (@) character.
<[S-carboxamidomethyl-L-cysteine]@C>ATPEILTCNSIGCLK
<[MOD:01090]@C>ATPEILTCNSIGCLK

2- Ambiguity in the modification position:
2.1- Unknown modification positions can be indicated with the sign '?'.

[Phospho]?EM[Oxidation]EVTSESPEK
[Phospho][Phospho]?[Acetyl]-EM[Oxidation]EVTSESPEK

2.2- Groups of modifications can be linked using arbitrary labels. The preferred location for the modification is indicated by the actual position of the modification tag or name in the amino acid sequence. Scores on the modification position can be indicated using parentheses. Any annotation made with the symbol # represents a way of linking different locations within the amino acid sequence. In ProForma 2.0 it is used for representing cross-linkers, branched peptides and for grouping protein modifications (including glycans) to represent ambiguity.

EM[Oxidation]EVT[#g1]S[#g1]ES[Phospho#g1]PEK

EM[Oxidation]EVT[#g1(0.01)]S[#g1(0.09)]ES[Phospho#g1(0.90)]PEK

2.3- The cases reported in 6.1 and 6.2 can be combined to represent scores of the modification position.

[Phospho#s1]?EM[Oxidation]EVT[#s1(0.01)]S[#s1(0.90)]ES[#s1(0.90)]PEK

2.4- A range of positions for a modification can be indicated in the amino acid sequence using a parenthesis for those amino acids involved.

PROT(EOSFORMS)[+19.0523]ISK
PROT(EOC[Carbamidomethyl]FORMS)[+19.0523]ISK

3- Chemical formulas of small molecules may be specified using the descriptor [Formula:].
3.1- A formula will be composed of pairs of atoms and their corresponding cardinality. Pairs MAY be separated by spaces.

SEQUEN[Formula:C12H20O2]CE

3.2- Cardinalities must be a positive or negative integer values. Zero is not supported. If a cardinality is not included with an atom, it is assumed to be +1.

SEQUEN[Formula:HN-1O2]CE

3.3- Isotopes will be handled by prefixing the atom with its isotopic number in square brackets.

Here, 2 $^{12}$C atoms are replaced by 2 $^{13}$C atoms:

SEQUEN[Formula:[13C2][12C-2]H2N]CE     (here 2 $^{12}$C atoms are replaced by 2 $^{13}$C atoms)

4- Glycan residues (generic monosaccharides) can be represented using the descriptor "Glycan"[Glycan:].

SEQUEN[Glycan:HexNAc]CE

5- Cross-Linked Peptides and branched peptides

Using the XL-MOD CV, crosslinked sites MUST be represented immediately after the modification notation using the prefix #xl, followed by an arbitrary label consisting of alphanumeric characters ([A-Za-z0-9]+ in regular expression notation). Cross-linker modification notations MUST be mentioned once only. As mentioned above, any annotation made with the symbol # represents a way of linking different locations within the amino acid sequence. In ProForma 2.0 it is used for representing cross-linkers, and for grouping protein modifications (including glycans) to represent ambiguity.

EMEVTK[XLMOD:02001#XL1]SESPEK[#XL1]
"Dead end" crosslink: EMEVTK[XLMOD:02001#XL1]SESPEK

Inter-protein or inter-chain connections are supported using // to separate the crosslinked peptides.

SEK[XLMOD:02001#XL1]UENCE//EMEVTK[XLMOD:02001#XL1]SESPEK

Branched peptides can be expressed using the same notation used for representing two cross-linked peptides, but using the term #BRANCH.

a) ETFGD[MOD:00093#BRANCH]//R[#BRANCH]ATER

6- Representation of ambiguity in the amino acid sequence. This mechanism can be used to represent changes that do not change the mass of the peptidoform/proteoform, but are not known. The way to encode this information is to use a parenthesis and a quotation mark including the ambiguous sequence represented in a preferred way.

(?DQ)NGTWEM[Oxidation]ESNENFEGYM[Oxidation]K

7- Joint representation of experimental data and its interpretation uses the pipe "|" character.

ELVIS[Phospho|+79.966331]K
ELVIS[Phospho|Obs:+79.978]K

8- INFO Tag. The information represented in between an INFO tag is considered non-standard (e.g. any text except a close bracket character) and does not need to be parsed. It is equivalent to a #comment in source code. They can be split using the pipe character.

ELV[INFO:xxxxx]IS
ELVIS[Phospho|INFO:newly discovered|INFO:really awesome]K

"Info" tags can be used to provide metadata about the ProForma entities.

ELVIS[Phospho|INFO:newly discovered|INFO:Created on 2021-06]K
ELVIS[Phospho|INFO:newly discovered|INFO:Created by software Tool1]K

9- Representation of isotopes: They can be represented using <> including the concrete isotope in between.

<13C>ATPEILTVNSIGQLK

10- Representation of mass spectra features:
10.1- Charges for spectra are indicated at the end of the sequence using /.

EMEVEESPEK/2

10.2 Chimeric spectra are indicated using the plus "+" character.

EMEVEESPEK+ELVISLIVER
EMEVEESPEK/2+ELVISLIVER/3

## 9. Authors Information

Juan Antonio Vizcaíno
European Bioinformatics Institute (EMBL-EBI), Hinxton, Cambridge, United Kingdom
juan@ebi.ac.uk

Eric W. Deutsch
Institute for Systems Biology, Seattle WA, USA
edeutsch@systemsbiology.org

Pierre-Alain Binz
CHUV Lausanne University Hospital, Lausanne, Switzerland
Pierre-Alain.Binz@chuv.ch

Ryan Fellers
Northwestern University, Evanston IL, USA
ryan.fellers@northwestern.edu

Anthony J. Cesnik
Stanford University, Stanford CA, USA
cesnik@stanford.edu

Joshua A. Klein
Boston University, Boston MA, USA
joshua.adam.klein@gmail.com

Tim Van Den Bossche
Ghent University, Ghent, Belgium; VIB-UGent Center for Medical Biotechnology, VIB, Ghent, Belgium
Tim.VanDenBossche@ugent.be

Ralf Gabriels
Ghent University, Ghent, Belgium; VIB-UGent Center for Medical Biotechnology, VIB, Ghent, Belgium
ralf.gabriels@ugent.be

Yasset Perez-Riverol
European Bioinformatics Institute (EMBL-EBI), Hinxton, Cambridge, United Kingdom
yperez@ebi.ac.uk

Jeremy Carver
University of California San Diego, San Diego CA, USA
jcarver@ucsd.edu

Shin Kawano
Toyama University of International Studies, Toyama, Japan

kawano@tuins.ac.jp

Benjamin Pullman
University of California San Diego, San Diego CA, USA
bpullman@eng.ucsd.edu

Nuno Bandeira
University of California San Diego, San Diego CA, USA
bandeira@ucsd.edu

Paul M. Thomas
Northwestern University, Evanston IL, USA
paul-thomas@northwestern.edu

Richard Leduc
Northwestern University, Evanston IL, USA
richard.leduc@northwestern.edu

## 10.     Contributors

In addition to the authors, a number of additional contributions have been made during the preparation process. The contributors who actively participated to the recommendation documentation are:

- Brian L. Frey, University of Wisconsin-Madison, Madison, WI, USA
- Alexander Leitner, ETH Zurich, Zurich, Switzerland
- Luis Mendoza, Institute for Systems Biology, Seattle, WA, USA
- Gerben Menschaert, Ghent University, Ghent, Belgium
- Jim Shofstahl, Thermo Fisher Scientific, San Jose, CA, USA
- Zhi Sun, Institute for Systems Biology, Seattle, WA, USA
- Leah V. Schaffer, University of Wisconsin-Madison, Madison, WI, USA
- Michael R. Shortreed, University of Wisconsin-Madison, Madison, WI, USA
- Veit Schwämmle, University of Southern Denmark, Odense, Denmark
- Wout Bittremieux, University of California San Diego, San Diego CA, USA

We would also like to acknowledge the contributions of the peer reviewers that participated in the PSI review process:
- Gloria Sheynkman and Erin Jeffery (University of Virginia, VA, USA).
- Xiaowen Liu (Tulane University School of Medicine, LA, USA).

Finally, we are also appreciative of the contributions made by the Executive Board of the CDTP, led by Paul Danis.

## 11.    Intellectual Property Statement

The PSI takes no position regarding the validity or scope of any intellectual property or other rights that might be claimed to pertain to the implementation or use of the technology described in this document or the extent to which any license under such rights might or might not be available; neither does it represent that it has made any effort to identify any such rights. Copies of claims of rights made available for publication and any assurances of licenses to be made available, or the result of an attempt made to obtain a general license or permission for the use of such proprietary rights by implementers or users of this specification can be obtained from the PSI Chair.

The PSI invites any interested party to bring to its attention any copyrights, patents or patent applications, or other proprietary rights which may cover technology that may be required to practice this recommendation. Please address the information to the PSI Chair (see contacts information at PSI website).

## 12.    Copyright Notice

http://psidev.info/proforma

### 13.    Glossary

All non-standard terms are already defined in detail in section 3.